\documentclass[a4paper,11pt,twoside,reqno]{amsart}

\usepackage{latexsym}
\usepackage[english]{babel}
\usepackage{fancyhdr}
\usepackage[mathscr]{eucal}
\usepackage[dvips]{graphicx}
\usepackage{graphics}
\usepackage{color}
\usepackage{epsfig}
\usepackage{amsmath}
\usepackage{amsfonts}
\usepackage{amssymb}
\usepackage{amscd}
\usepackage{bbm}
\usepackage{latexsym}
\usepackage{pifont}
\usepackage{subfigure}
\usepackage{psfrag}
\usepackage{caption}
\newcommand{\ud}{\mathrm{d}}

\newcommand{\re}{\mathfrak{Re}}
\newcommand{\im}{\mathfrak{Im}}

\theoremstyle{definition} 
\theoremstyle{definition}
\theoremstyle{definition}

\numberwithin{equation}{section}

\renewcommand\thepage{}
\begin{document}

\thispagestyle{empty}

\Large

\textbf{O.~Zagordi and A.~Michelangeli}

\vspace{2cm}

\textbf{1D PERIODIC POTENTIALS} 

\textbf{WITH GAPS VANISHING AT $k=0$}

\normalsize

\newpage

\clearpage

\thispagestyle{empty}

\textbf{Abstract.} Appearance of energy bands and gaps in the dispersion relations of a periodic potential is a standard feature of Quantum Mechanics. We investigate the class of one-dimensional periodic potentials for which all gaps vanish at the center of the Brillouin zone. We characterize them through a necessary and sufficient condition. Potentials of the form we focus on arise in
different fields of Physics, from supersymmetric Quantum Mechanics, to Korteweg-de Vries equation theory and classical diffusion problems. 
The O.D.E.~counterpart to this problem is the characterisation of
periodic potentials for which coexistence occur of linearly independent solutions of the corresponding Schr\"odinger equation (Hill's equation).
This result is placed in perspective of the previous related results available in the literature.

\medskip

\textbf{2000 Mathematics Subjects Classification.} 30RD10, 30RD15, 30RD20, 34B24, 34B30, 34E05, 34L05, 34L99, 46N20, 46N50, 47N20, 47N50, 81Q10, 81V45


\medskip

 Intervals of stability and instability. Discriminant and characteristic values of an O.D.E. Coexistence.

\newpage

\section{Introduction}

A well-known achievement of Quantum Mechanics is the understanding of the band structure of the energy spectrum for periodic potentials \cite{ash}. Since ever the characterisation of dispersion relations and gaps between bands of permitted energy has turned out to be of importance, due to their crucial role in the conductor-insulator properties of crystalline solids.

In this scenario, the case of one-dimensional (1D) periodic potentials, despite its simplicity, is physically meaningful not only for pedagogical reasons, but also because it models real structures with a preferred direction (as nanotubes or nanowires).

Our interest here is to study a sub-class of the very general problem of `\emph{vanishing gaps}', focusing on real 1D periodic potentials of the form $W^2+W'+v_0$, $v_0$ being a constant and $W(x)$ changing sign after half a period. These potentials -- and only these -- turn out to have \emph{all} the gaps vanishing at the centre of the Brillouin zone.

Remarkably, one ends up with a (not necessarily periodic) 1D potential of the form $v(x)=W(x)^2+W'(x)$ in several different fields of Physics,
as in supersymmetric Quantum Mechanics \cite{cks95}, where spectral analysis combined with the formalism of SuSYQM has led to a large class of analytically
solvable 1D periodic potentials \cite{ks99,klgy01,ks04jphys}, and peculiar features of supersymmetry breaking can be exploited \cite{dm98,dfein98}.
Other examples are in the framework of the Korteweg-de Vries equation, via the Miura transform \cite{miura,calogero}, and in the mapping of a Fokker-Plank equation
onto a quantum stationary problem \cite{mr79,risken}, together with simulation methods related \cite{baronimoroni98}.

The outline of this paper is as follows. In Section \ref{scenario} we state the physical problem and its
mathematical formulation.
Also, we mention some previous results on the criteria which make
energy gaps disappear. In Section \ref{sect3} we state and discuss our vanishing-gaps results with some examples.
Section \ref{proofs} shows the proofs by means of elementary Quantum Mechanics and operator theory,
as well as standard O.D.E. and Complex Analysis theory.
The appendix contains a more detailed review of the mathematics underlying the band structure theory.

\section {Background Theory}\label{scenario}

A Schr\"odinger-like particle in a periodic 1D potential is described by the Hamiltonian
\begin{equation}\label{Hamiltonian}
H=-\frac{\ud^2}{\ud x^2}+v(x),
\end{equation}
where $v(x+1)=v(x)$. A unit period for $v$, as well as units $\hbar=2m=1$, are assumed without loss of generality. We take $v$ to be real and suitably regular, as specified later. This means that the first Brillouin zone is bounded by $-\pi,\pi$.

According to the Floquet-Bloch theory, the eigenfunctions of $H$ can be chosen of the form $\psi_k(x)=e^{ikx}u_k(x)$, with $u_k(x+1)=u_k(x)$, namely a plane wave $e^{ikx}$ modulated by a periodic function $u_k$ of the same periodicity of $v$. The energy $E(k)$ of the eigenfunction $\psi_k$, when plotted against $k$, gives the well-known band structure. Forbidden and permitted energies, bands filling, conductivity, and other related features can then be discussed in view of such dispersion relations \cite{ash,kittel,gp2000}.

Since we are interested in the vanishing of the gaps at the centre ($k=0$) and at the edge ($k=\pi$) of the Brillouin zone, we study the Hamiltonian (\ref{Hamiltonian}) as the operator $H(0)$ or $H(\pi)$ defined on the domain $\mathcal{D}_+$ and $\mathcal{D}_-$ respectively, of the measurable functions on the interval $[0,1]$ that are square-summable together with their first two  derivatives, and satisfy the boundary conditions
\begin{equation}\label{periodicity}
\begin{array}{lllllllll}
\psi(1)&=&\phantom{-}\psi(0)\, ,&\quad&\psi'(1)&=&\phantom{-}\psi'(0)&\;\;&(\mathcal{D}_+)\\
\psi(1)&=&          -\psi(0)\,,& &\psi'(1)&=&          -\psi'(0)&\quad&(\mathcal{D}_-)\,.
\end{array}
\end{equation}

\begin{figure}[ht]
\begin{center}
\includegraphics[height=5cm]{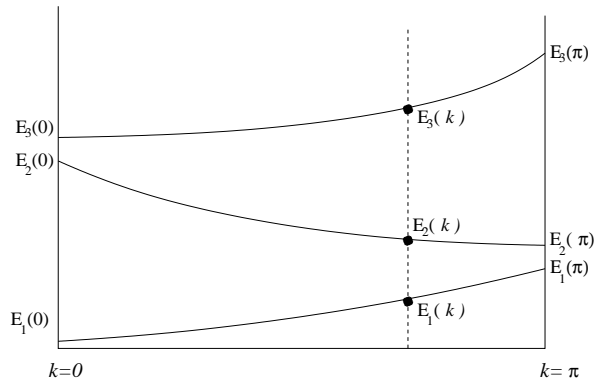}
\end{center}
\caption{Typical dispersion relations for a 1D periodic potential}\label{bande}
\end{figure}

One recovers the Bloch periodic/antiperiodic functions simply extending any $\psi\in\mathcal{D}_\pm$ by periodicity on $\mathbb{R}$, and the familiar structure with bands and gaps as depicted in Fig.~\ref{bande} with the customary notations. Eigenfunctions at $k=0$ have period equal to $1$, while those at $k=\pi$ have period equal to $2$. One band and its subsequent collapse into a unique band in the centre of the Brillouin zone at the energy $E$, iff $E$ is a \emph{doubly degenerate} eigenvalue of $H(0)$. Similarly they collapse at the edge of the Brillouin zone iff $E$ is a \emph{doubly degenerate} eigenvalue of $H(\pi)$.

A more detailed description of this theory, which in the mathematical literature is known as the
\emph{Hill's equation} theory, is reported in the appendix.

Notice that the emergence of vanishing gaps is quite `unusual', as for most potentials all gaps have a nonzero width \cite{mckean76,simon76}.
In fact, in a sense that resembles the way that real numbers can be suitably approximated by
rationals, any potential with vanishing gaps can be suitably approximated with a potential with all nonzero gaps.
This establishes a sort of `peculiarity condition' for potentials with vanishing gaps,
which motivates the interest towards them.

An example of 1D periodic potential which is well known to exhibit some vanishing gaps in the dispersion relations is
the square well \emph{Kronig-Penney model} \cite{kp31} when certain conditions on the depth and length of the
wells are met \cite{ls79,fol83}. Nevertheless, infinitely many nonzero gaps always occur.

Through numerical analysis vanishing gaps have been found to occur for a hybrid of triangular wells separated by flat
interstitial regions \cite{ls79}. The emergence of vanishing gaps has been established
by perturbative analysis on a smooth potential \cite{ml92}, dispelling the misleading idea that in some way only flat
sections are associated with the vanishing phenomenon.

The \emph{Whittaker-Hill potential}, also known as the \emph{trigonometric Ra\-zavy potential}
\cite{ks04jphys,magnus,r79,km98,fgr99}, is an example where all gaps except for a finite number of them,
vanish only at the centre or only at the edge of the Brillouin zone.
In the particular case $n=0$ we have the example given in equation (\ref{our_example}) below.

Only a finite number of nonzero gaps appear in the class of the \emph{Lam\'e}
and \emph{associated Lam\'e potentials} \cite{ks99,ks04jphys,magnus,s2001,ks01,ks04physlett}: depending on the
choice of the parameters entering their definition, they still exhibit a finite number of bound bands followed by an infinite continuum band, and a well-studied
pattern of vanishing gaps. Analogous results have been proved to hold for some complex-valued PT-invariant versions
\cite{ks04jphys,ks04physlett,ks05}.

Beyond these examples, on the other hand, necessary conditions are known on a 1D periodic potential $v$ with a prescribed
number of nonzero gaps in its dispersion relations. In particular \cite{rs4_in}:
\begin{enumerate}
\item if \emph{no} gaps are present, then $v$ is a constant \cite{borg46,ungar61,hoch63,hoch65};
\item if precisely one gap occur, $v$ is a Weierstrass elliptic function \cite{hoch65,dub74,hoch76};
\item if only finitely many gaps are present, then $v$ is real analytic as a function on the reals \cite{mckean76,gol74,g81};
\item if $v(x+a)=v(x)$ and all gaps at $k=\pi$ are absent, then $v(x+\frac{a}{2})=v(x)$ \cite{borg46,hoch66,h77}.
\end{enumerate}

In view of (4) above, we may assume that $v$ has period 1 in the sense that
\[
1=\min\{a>0\,:\,v(x+a)=v(x)\}
\]
(incidentally, this excludes $v$ to be trivially a constant); if so, some gap at $k=\pi$ must be open, whereas nothing is said at $k=0$. It is this question that we are facing in the following.

\section{Real periodic potentials $v_0+W(x)^2+W'(x)$ \\ with $W(x+\frac{1}{2})=-W(x)$}\label{sect3}

We now come to the main object of our analysis.

\noindent\textbf{Theorem.} \emph{Let $v$ be a 1-periodic continuous potential. A necessary and sufficient condition for $v$
to have \emph{all} gaps vanish at the centre of the Brillouin zone is that
\begin{equation}\label{TH}
v(x)=v_0+W^2(x)+W'(x)
\end{equation}
for some constant $v_0$ and some differentiable $W$ changing sign after half a period:
\begin{equation}\label{changesign}
W(x+\frac{1}{2})=-W(x)\,.
\end{equation}
Moreover, in terms of $v$, the function $W$ is given by
\begin{equation}\label{W}
W(x)=-\frac{1}{2}\int_x^{x+\frac{1}{2}}\!\Big[ v(\xi)-\int_0^1 v(\zeta)\,\ud\zeta\Big]\,\ud\xi\,,
\end{equation}
thus it is determined only by the odd harmonic part of $v$:
\begin{equation}\label{W'}
W'(x)=\frac{v(x)-v(x+\frac{1}{2})}{2}\,,
\end{equation}
whereas the constant $v_0$ is given by
\begin{equation}\label{v0}
v_0=\int_0^1\big[v(x)-W^2(x)\big]\,\ud x\,.
\end{equation}
Conditions (\ref{TH}) and (\ref{changesign}) uniquely fix $W$ and hence $v_0$ to have the form (\ref{W}) and (\ref{v0})
respectively. Also, whenever (\ref{TH}) and (\ref{changesign}) hold, then $v_0$ is the lowest energy in the dispersion
relations of $v$, i.e., it is the ground state of the Hamiltonian $H(0)$ with periodic boundary conditions}.

In other words, the theorem states that given a 1D periodic potential $v$ (with period 1) and $W$ and $v_0$ as in (\ref{W})
and (\ref{v0}) respectively, then $v-[v_0+W^2+W']\equiv0$ if and only if all the gaps vanish at the centre of the Brillouin
zone. In particular, when $W$ is given as in (\ref{W}), $W^2+W'$ has a zero-energy ground state, and $v_0+W^2+W'$ has ground
state $v_0$. Thus, if 
$E_{GS}$ is the ground state of any 1D 1-periodic $v$, 
condition $E_{GS}-v_0\neq 0$ necessarily
implies that at least one gap exists at the centre of the Brillouin zone and the quantity $E_{GS}-v_0$ can be seen
as a `measure' of such a non-vanishing phenomenon.

We mention that a more detailed analysis\footnote{same authors, in preparation} shows that with a bit
of standard (although non trivial) functional-analytic technicalities, the regularity of $W$ can be considerably weakened.

It is worth noticing that potentials characterised by the theorem above may or may not have some vanishing gaps at the
\emph{edge} of the Brillouin zone as well; nevertheless, as stated above, some of them must be necessarily nonzero, unless
the potential is a constant.

\begin{figure*}
\begin{center}
\includegraphics[height=5.5cm]{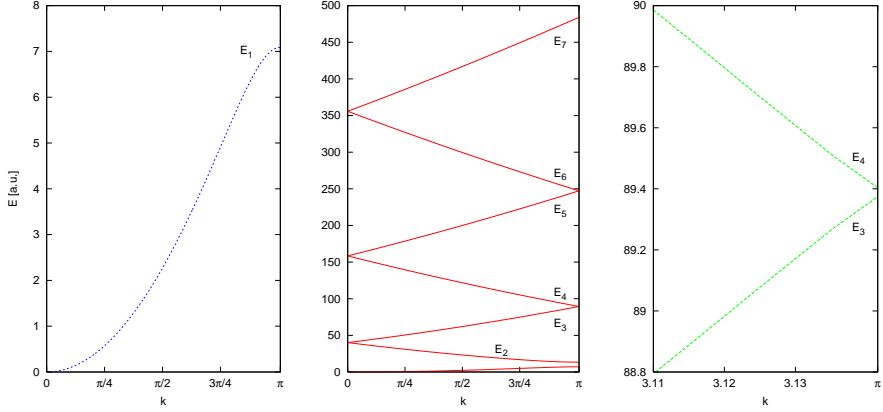}
\end{center}
\caption{Centre: band structure for $v(x)=2\pi\cos 2\pi x + \sin^2 2\pi x$. Left: close-up of the lowest energy band. Right:
at $k=\pi$ bands do not overlap.}\label{seno}
\end{figure*}

As an example let us concentrate on the very simple choice $W(x)=\sin 2\pi x$. The corresponding potential
\begin{equation}\label{our_example}
v(x)=W^2(x)+W'(x)=2\pi\cos 2\pi x + \sin^2 2\pi x
\end{equation}
has period 1, and is a particular case of the Razavy potential we mentioned above. According to the theorem all gaps vanish
at the centre of the Brillouin zone and the ground state is zero. The dispersion relations for $v$ are plotted in Fig.~\ref{seno}.
The emergence of nonzero gaps at the edge of the Brillouin zone is necessarily expected, our $v$ being not a constant: indeed
some open gaps at $k=\pi$ are clearly visible. We underline that this very simple potential has only two Fourier components,
making it interesting also in the field of optical lattices.

\begin{figure}
\begin{center}
\includegraphics[height=6cm]{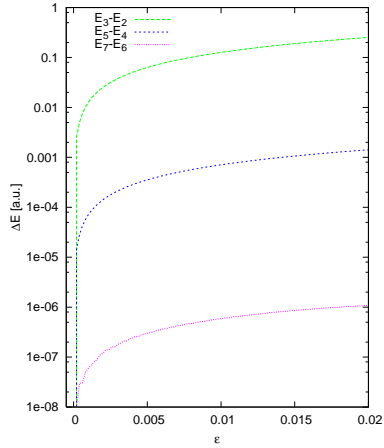}
\end{center}
\caption{Removal of the $k=0$ vanishing gaps for the potential $v$ corresponding to the perturbation
$W(x)=\sin 2\pi x \mapsto\sin 2\pi x +\varepsilon\cdot\sin 4\pi x$}\label{sevarioepsilon}
\end{figure}

When we slightly perturb $W(x)=\sin 2\pi x$ in such a way that the condition (\ref{changesign}) is destroyed, e.g.,
by substituting
\begin{equation}
W(x)\longmapsto W(x)+\varepsilon\eta(x)
\end{equation}
where $\eta(x+\frac{1}{2})\equiv\!\!\!\!\!/ -\eta(x)$,
the doubly degenerate eigenvalues at $k=0$ split, and non-vanishing gaps appear separating the bands.
Such a behaviour is reproduced in Fig.~\ref{sevarioepsilon}.

As another example we point out the \emph{one-gap} Lam\'e potentials, which turn out to be a subclass of those we are dealing
with, in that they show a single gap (the first one) at the edge of the Brillouin zone, whereas all gaps vanish at the centre.
They have recently received new attention \cite{fromhold} since they optimise some key parameters in the band structure
and are of practical interest in the realization of quasi one-dimensional crystals.

To summarise, we have seen that \emph{the entire class} of potentials with all gaps vanishing at $k=0$ is analytically characterised by
the very simple formulas (\ref{TH}) and (\ref{W}), while the typical issue one can find in the literature is to identify \emph{specific} potentials
with a given pattern of open and vanishing gaps (Kronig-Penney, Razavy, Lam\'e, \dots). Nevertheless, the formulas
characterising this family might be used for band design purposes, or conversely to check whether a given potential belongs
to this class.

\section{Proofs}\label{proofs}

\subsection{Proof of sufficiency}

We want to make use of elementary operator theory, unlike the usual O.D.E. approach one can
find in the literature (see appendix). For convenience let us rename the Hamiltonian $H(0)$, as defined in (\ref{H(k)}), as $H:=H(0)$.
Once we set $v_0=0$ the claim amounts to say that $H$ is a positive operator, that its ground state is $E_1=0$, and that any other
level $E>0$ is doubly degenerate. Also, recall by (\ref{periodicity}) and (\ref{k-periodicity}) that the domain $\mathcal{D}_+$ of $H$
is characterised by the periodic boundary conditions
\begin{equation}
\psi(1)=\psi(0),\quad\psi'(1)=\psi'(0)\,.
\end{equation}

By means of the operator
\begin{equation}
a :=\frac{\ud}{\ud x}-W(x)
\end{equation}
and its adjoint
\begin{equation}
a^\dagger=-\frac{\ud}{\ud x}-W(x)
\end{equation}
the Hamiltonian can be factorised as
\begin{equation}\label{a^*a}
H=a^\dag a.
\end{equation}
Notice that $W$ is regular enough to guarantee that (\ref{a^*a}) makes sense without domain problems: both $a$ and $a^\dag$ are only
densely defined in $L^2[0,1]$, nevertheless since $W$ is differentiable, $a$ maps $\mathcal{D}_+$ \emph{into} the domain of $a^\dag$,
so that $a^\dag a$ makes sense on the whole $\mathcal{D}_+$ (just as $a\,a^\dag$ does).

Of course factorisation (\ref{a^*a}) means that $H$ is a positive operator, hence all its eigenvalues are nonnegative.
By direct inspection one can check that the ground state is $E_1=0$: indeed the (non normalised) state 
\begin{equation}
\psi_1(x)=e^{\int^x_0 W(\xi)\,\ud \xi}
\end{equation}
is annihilated by $a$, i.e., $a\,\psi_1=0$, since it solves the O.D.E.
\begin{equation}
\psi'(x)-W(x)\psi(x)=0
\end{equation}
and then also $H\psi_1=0$ holds true. $E_1$ is always nondegenerate, as we recall in the appendix.
Notice, in addition, that the ground state wave function $\psi_1$ can be chosen to be \emph{real}.

We now want to exploit another factorisation, beyond the standard one (\ref{a^*a}). Let $T$ be the operator of translation of half a period
\begin{equation}
T\psi(x)=\psi(x+\frac{1}{2}),
\end{equation}
defined with the natural periodicity in order that it becomes a unitary transformation on $L^2[0,1]$, i.e.,
\begin{equation}
TT^\dagger=T^\dagger T=\mathbbm{1}\,.
\end{equation}

As a consequence of the assumption (\ref{changesign}), the other operators transform under $T$ as
\begin{equation}
\begin{array}{lllllll}
TWT^\dagger & = & -W\,,         & \quad & T\displaystyle{\frac{\ud}{\ud x}}T^\dagger & = & \displaystyle{\frac{\ud}{\ud x}}\,, \\
TaT^\dagger & = & -a^\dagger\,, & \quad & Ta^\dagger T^\dagger        & = & -a\,,
\end{array}
\end{equation}

\begin{equation}
\Rightarrow\quad THT^\dagger=T(a^\dagger a)T^\dagger=a\,a^\dagger\,, \label{underT}
\end{equation}
so that one can conveniently rewrite
\begin{equation}\label{factorization}
H\;=\;a^\dag a\;=\;(Ta)^\dag(Ta)\;=\;(Ta)(Ta)^\dag
\end{equation}
and get the commutation relations
\begin{equation}\label{commutativity}
[Ta,(Ta)^\dag]\;=\;\left[H,Ta\right]\;=\;\left[H,(Ta)^\dag\right]\;=\;\mathbb{O}\,.
\end{equation}
It follows that any energy level is invariant under the action of $Ta$ or $(Ta)^\dag$.

On $\mathcal{D}_+$ obviously $T$ and $T^\dag$ square to unity; moreover $Ta$ is anti-hermitian, i.e., $(Ta)^\dagger=-Ta$. Indeed, for any $\psi\in\mathcal{D}_+$ and any $x\in[0,1]$ 
\begin{equation}\label{antiermitean1}
\begin{array}{lll}
(Ta)^\dagger \psi(x)&=&a^\dag T^\dag\psi(x) \\
&=&-\psi'(x-\frac{1}{2})-W(x)\psi(x-\frac{1}{2}) \\
&=&-\Big[\,\psi'(x-\frac{1}{2})+W(x)\psi(x-\frac{1}{2})\,\Big] \\
&=&-\Big[\,\psi'(x+\frac{1}{2})-W(x+\frac{1}{2})\psi(x+\frac{1}{2})\,\Big] \\
&=&-(Ta)\psi(x)\,.
\end{array}\end{equation}
The anti-hermiticity of $Ta$ implies that all its nonzero eigenvalues are pure imaginary numbers and that eigenfunctions belonging to distinct eigenvalues are orthogonal. 

Take now any $\psi\in\mathcal{D}_+$ and any $\lambda\neq 0$ in $i\mathbb{R}$ such that $Ta\psi=\lambda\psi$. Then $H\psi=|\lambda|^2\psi$, because $H=(Ta)^\dag(Ta)=-(Ta)^2$, that is, $\psi$ has energy $E=|\lambda|^2$. Taking all such possible choices of $\psi$'s and $\lambda$'s, one recovers \emph{all} the nonzero energy eigenvalues, because of the commutation relations (\ref{commutativity}).

So, let $Ta\psi=\lambda\psi$:
\begin{equation}
\psi'(x+\frac{1}{2})-W(x+\frac{1}{2})\psi(x+\frac{1}{2})=\lambda\psi(x)\,;
\end{equation}
by conjugation, since $W(x)\in\mathbb{R}$ and $\lambda\in i\mathbb{R}$, this is equivalent to
\begin{equation}
\bar\psi'(x+\frac{1}{2})-W(x+\frac{1}{2})\bar\psi(x+\frac{1}{2})=\bar\lambda\bar\psi(x)=-\lambda\bar\psi(x)\,,
\end{equation}
$\bar\psi(x)$ being the complex conjugate of $\psi(x)$.
That is, $Ta\bar\psi=-\lambda\bar\psi$. Both $\psi$ and $\bar\psi$ have energy $E=|\lambda|^2$, but they are orthogonal, since $\lambda\neq 0\Rightarrow\lambda\neq-\lambda$ (distinct eigenvalues of $Ta$). Therefore, one has shown that the energy level $E$ is doubly degenerate,
and the conclusion must hold for \emph{any} $E>0$. $\hspace{\stretch{1}}\square$

Notice that conjugating the eigenvalue problem $Ta\psi=\lambda\psi$ turns out to be useful because it leads to the orthogonality of $\psi$ and $\bar\psi$. The same does not apply to the eigenvalue problem for $H$, in that $H\psi=E\psi$ do imply $H\bar\psi=E\bar\psi$, but one cannot argue that $\psi$ and $\bar\psi$ are linearly independent.

Concerning the possibility that some gaps (or all of them) vanish at the edge $k=\pi$ of the Brillouin zone,
it is clear that the scheme of the proof above does \emph{not} apply.
Indeed, at $k=\pi$ the spectral analysis has to be performed on the Hamiltonian $H(\pi)$ now defined on the domain
$\mathcal{D}_-$ (\emph{antiperiodic} boundary conditions); although the same factorisation (\ref{factorization})
in terms of $Ta$ still holds, when one tries to mimic (\ref{antiermitean1}) one now gets
\begin{equation}\begin{array}{lll}
(Ta)^\dagger \psi(x)&=&a^\dag T^\dag\psi(x) \\
&=&-\psi'(x-\frac{1}{2})-W(x)\psi(x-\frac{1}{2}) \\
&=&\psi'(x+\frac{1}{2})-W(x+\frac{1}{2})\psi(x+\frac{1}{2}) \\
&=&(Ta)\psi(x)
\end{array}\end{equation}
that is, $Ta$ is \emph{hermitian} on the antiperiodic functions. Then its eigenvalues are real numbers and conjugating $Ta\psi=\lambda\psi$ one gets  $Ta\bar\psi=\lambda\bar\psi$ without being able to conclude whether $\psi\bot\bar\psi$ or not.

\subsection{Proof of necessity}\label{proof_of_necessity}

Let us assume for the moment that $v$ has mean zero:
$v_{\mathrm{mean}}:=\int_0^1v(x)\,\ud x=0$. In the notation of the appendix, let $\psi_1^{[E]}$ and $\psi_2^{[E]}$ be the
fundamental solutions and let $D(E)$ be the discriminant of $H\psi=E\psi$, $H$ being the Hamiltonian $H(0)$ of the proof
above, and of (\ref{H(k)}), and $E$ being any complex number. 

The object of crucial interest in this proof turns out
to be the function
\begin{equation}\label{entire1}
f(E)\equiv\frac{\psi_2^{[E]}(\frac{1}{2})-\psi_2^{[E]}(-\frac{1}{2})}{\sqrt{\frac{2-D(E)}{E-E_0}}}\,.
\end{equation}
In fact the assumption of all gaps vanishing at $k=0$ translates into
\begin{equation}
2-D(E)=\mathcal{M}^2(E)(E-E_0)
\end{equation}
where all the zeroes of $\mathcal{M}(E)$ are simple and $E_0$ is the lowest real root of $D(E)=2$. 
Now, let $\mathcal{G}^{[E]}(x)$ be the Green's function of the problem
\begin{equation}\label{Green}
\Big[-\frac{\ud^2}{\ud x^2}+v(x)-E\Big]\,\mathcal{G}^{[E]}(x)+\delta(x)=0
\end{equation}
with periodic boundary conditions on $[0,1]$; 
by means of standard O.D.E. techniques, it is seen to be
\begin{equation}
\mathcal{G}^{[E]}(x)=\frac{\psi_2^{[E]}(x)-\psi_2^{[E]}(x-1)}{2-D(E)}\,.
\end{equation}
Evaluating $\mathcal{G}^{[E]}$ at mid-period ($x=\frac{1}{2}$) one gets
\begin{equation}
\mathcal{G}^{[E]}(\frac{1}{2})=\frac{\psi_2^{[E]}(\frac{1}{2})-\psi_2^{[E]}(-\frac{1}{2})}{\mathcal{M}^2(E)(E-E_0)}\,
\end{equation}
and since the boundary value problem (\ref{Green}) is self-adjoint, such a Green's function can have only simple poles
in the energy complex plane. Accordingly, by multiplication by $\mathcal{M}(E)(E-E_0)$
\begin{equation}\begin{split}
\mathcal{G}^{[E]}(\frac{1}{2})\mathcal{M}(E)(E-E_0)&=\frac{\psi_2^{[E]}(\frac{1}{2})-\psi_2^{[E]}(-\frac{1}{2})}{\mathcal{M}(E)}\\
 &=\frac{\psi_2^{[E]}(\frac{1}{2})-\psi_2^{[E]}(-\frac{1}{2})}{\displaystyle\sqrt{\frac{2-D(E)}{E-E_0}}}=f(E)
\end{split}\end{equation} 
must be an \emph{entire} function. 

Our claim at this point is that $f$ is also bounded (as $E\to\infty$): then by Liouville's theorem it is identically constant.
We enclose such claim in the following somehow technical Lemma.
Its proof is postponed to the next subsection. Hereafter, the mainstream of the proof continues.

Denote by
\begin{eqnarray}
\mathcal{N}(E) & := & \psi_2^{[E]}(\frac{1}{2})-\psi_2^{[E]}(-\frac{1}{2}) \\
\mathcal{M}(E) & := & \displaystyle\sqrt{\frac{2-D(E)}{E-E_0}}
\end{eqnarray}
the numerator and the denominator of $f$ respectively.

\noindent\textbf{Lemma (uniform boundedness of $f$ and estimates).} \emph{The entire function $f^2$, and hence $f$, is uniformly bounded at infinity. Consequently it is identically constant, such a constant being 1 due to a direct evaluation. 
Further, in the region}
\begin{equation}\label{omegaepsilon}
\Omega_{\varepsilon}\equiv\{E\in\mathbb{C}\;:\;\varepsilon \leqslant\arg(E)\leqslant 2\pi-\varepsilon\}
\end{equation}
\emph{($\varepsilon$ being any fixed small strictly positive number)  the following asymptotic estimates hold for the square of the numerator}
\begin{equation}\label{numerator}\begin{split}
\mathcal{N}^2(E) & = \displaystyle\frac{4\sin^2\frac{\sqrt{E}}{2}}{E}+c\,\frac{\sin^2\frac{\sqrt{E}}{2}}{E^2}+\mathcal{O}\Big(\frac{e^{\im\sqrt{E}}}{E^{5/2}}\Big) \\
c & = 2\,\Big[\,v(\frac{1}{2})+v(0)-\Big(\int_0^{1/2}\!\!v(\xi)\,\ud\xi\Big)^2\,\Big]
\end{split}\end{equation}
\emph{and the square of the denominator}
\begin{equation}\label{denominator}
\mathcal{M}^2(E)=\displaystyle\frac{4\sin^2\frac{\sqrt{E}}{2}}{E}+4E_0\,\frac{\sin^2\frac{\sqrt{E}}{2}}{E^2}+\mathcal{O}\Big(\frac{e^{\im\sqrt{E}}}{E^{5/2}}\Big)
\end{equation}
\emph{where in both expansions each term is leading w.r.t.~the subsequent, as $E\to\infty$ in $\Omega_{\varepsilon}$.}

Thanks to this Lemma the proof of necessity is completed as follows. By comparison of the coefficients of the 
$\big(\frac{\sin^2\frac{\sqrt{E}}{2}}{E^2}\big)$--terms in (\ref{numerator}) and in (\ref{denominator}), one has $c=4E_0$, whence
\begin{equation}
2 E_0=v(\frac{1}{2})+v(0)-\frac{1}{2}\Big(\int_0^{\frac{1}{2}} v(\xi)\,\ud\xi\Big)^2\,.
\end{equation}
Since the ground state $E_0$ cannot change by any shift in $x$, this is the same as
\begin{equation}\begin{split}
2 E_0&=v(x+\frac{1}{2})+v(x)-\frac{1}{2}\Big(\int_x^{x+\frac{1}{2}} v(\xi)\,\ud\xi\Big)^2 \\ 
&=v(x+\frac{1}{2})+v(x)-\frac{1}{2}\Big(-2 W(x)\Big)^2 \,,
\end{split}\end{equation}
where $W$ is \emph{defined} by (\ref{W}) -- recall that we are now dealing with a zero-mean periodic $v$. Equivalently,
\begin{equation}
\frac{v(x)+v(x+\frac{1}{2})}{2}=E_0+W^2(x)\,.
\end{equation}
On the other side, definition (\ref{W}) clearly implies (\ref{W'}), namely
\begin{equation}
\frac{v(x)-v(x+\frac{1}{2})}{2}=W'(x)\,,
\end{equation}
so that altogether
\begin{equation}\begin{split}\label{thesis-zeromean}
v(x)&=\frac{v(x)+v(x+\frac{1}{2})}{2}+\frac{v(x)-v(x+\frac{1}{2})}{2} \\
&=E_0+W^2(x)+W'(x)\,.
\end{split}\end{equation}
The constant $E_0$ is recovered in terms of $W$ by integrating over one period and taking into account
that $\int_0^1v(x)\,\ud x=\int_0^1W'(x)\,\ud x=0$:
\begin{equation}\label{E_0}
E_0=-\int_0^1 W^2(x)\,\ud x\,.
\end{equation}

So far $v$ has had mean zero. For a generic $v$, the result above leads to $\tilde{v}=v-v_{\mathrm{mean}}=E_0+W^2+W'$, where $W$
is given by the full form of (\ref{W}).
Hence $v=v_0+W^2+W'$ with $v_0$ given by (\ref{v0}), and (\ref{TH}) is proved.

The uniqueness of $W$ is a consequence of the unique decomposition
$v=v_++v_-$, for any $1$-periodic function $v$, where $v_{\pm}(x+\frac{1}{2})=\pm v_{\pm}(x)$.
In fact, if $v=v_0+W^2+W'=u_0+U^2+U'$ for some constants $v_0$,
$u_0$ and some $W$, $U$ changing sign after half a period, then necessarily
\begin{equation}
\begin{array}{llclc}
v_+ &=& v_0 + W^2 &=& u_0 + U^2 \\
v_- &=& W'        &=& U'
\end{array}
\end{equation}
whence $U=W+const.$ and by substitution into $v$ the constant turns out to be zero. $\hspace{\stretch{1}}\square$

\subsection{Proof of the Lemma}\label{proof_of_lemma}

We proceed along the following steps.

\medskip

\noindent\textsc{Step 1.} \emph{$f^2$ is an entire function for which the following asymptotic estimate holds as $E\to\infty$}:
\begin{equation}\label{step1}
\!\!\!f^2(E)=\frac{\mathcal{N}^2(E)}{\mathcal{M}^2(E)}=\frac{\;\;\displaystyle\frac{4\sin^2\frac{\sqrt{E}}{2}}{E}+c\,\frac{\sin^2\frac{\sqrt{E}}{2}}{E^2}+\mathcal{O}\Big(\frac{e^{\im\sqrt{E}}}{E^{5/2}}\Big)\;}{\;\;\displaystyle\frac{4\sin^2\frac{\sqrt{E}}{2}}{E}+4E_0\,\frac{\sin^2\frac{\sqrt{E}}{2}}{E^2}+\mathcal{O}\Big(\frac{e^{\im\sqrt{E}}}{E^{5/2}}\Big)\;}
\end{equation}
\emph{with}
\begin{equation}
c=2\,\Big[\,v(\frac{1}{2})+v(0)-\Big(\int_0^{1/2}\!\!v(\xi)\,\ud\xi\Big)^2\,\Big]\,.
\end{equation}
Notice that by no means this suffices to say that the ratio is $\mathcal{O}(1)$: indeed the remainders are not meant to be necessarily subleading 
w.r.t.~the $E^{-1}\sin^2\frac{\sqrt{E}}{2}$ and the $E^{-2}\sin^2\frac{\sqrt{E}}{2}$--terms, since the latter vanish in the sequences of points $(2n\pi)^2$.


$f^2$ is entire because $f$ is. 
The rest of step 1 is simply a consequence of plugging into (\ref{entire1}) the following asymptotic estimates available in the literature \cite{magnus,hoch65}:
\begin{eqnarray}
2-D(E) & = & 4\sin^2\frac{\sqrt{E}}{2}+\mathcal{O}\Big(\frac{e^{\im\sqrt{E}}}{E^{3/2}}\Big)\label{2-D} \\
\psi_2^{[E]}(x) & = & \frac{\sin x\sqrt{E}}{\sqrt{E}} - \frac{\cos x\sqrt{E}}{2 E}\mathcal{V}(x) + \label{psi2} \\
& + & \frac{\sin x\sqrt{E}}{4 E^{3/2}}\,\big[\,v(x)+v(0)-\mathcal{V}^2(x)\,\big]+\mathcal{O}\Big(\frac{e^{\,x\,\im\frac{\sqrt{E}}{2}}}{E^2}\Big)\nonumber
\end{eqnarray}
($\forall E\in\mathbb{C}$ and uniformly in $x\in[0,1]$), where
\begin{equation}
\mathcal{V}(x):=\int_0^x v(\xi)\,\ud\xi\,.
\end{equation}
These are extensively discussed in the appendix. 
We only remark that the continuity of $v$ is all what is needed to truncate these expansions 
up to this order: to go higher to $\mathcal{O}(e^{\im\sqrt{E}}E^{-n/2})$-terms one has to assume $v$ to be 
sufficiently differentiable,
which is not needed here. 
Also, we stress again that, until a given complex path $E\to\infty$ is specified, 
it is not possible to identify the leading terms in (\ref{2-D}) and (\ref{psi2}).

From (\ref{psi2}) on gets
\begin{equation}\begin{split}
\mathcal{N}(E) &= \psi_2^{[E]}(\frac{1}{2})-\psi_2^{[E]}(-\frac{1}{2}) \\
& = \frac{2\sin\frac{\sqrt{E}}{2}}{\sqrt{E}}-\frac{2\cos\frac{\sqrt{E}}{2}}{2E}\big(\mathcal{V}(\frac{1}{2})-\mathcal{V}(-\frac{1}{2})\big) \\
& \;\;+\frac{\sin\frac{\sqrt{E}}{2}}{4 E^{3/2}}\,\Big[\,v(\frac{1}{2})+v(-\frac{1}{2})+2v(0)- \\
& \qquad\qquad\qquad - \frac{1}{2}\mathcal{V}^2(\frac{1}{2})-\frac{1}{2}\mathcal{V}^2(-\frac{1}{2})\,\Big] \\
& \;\;+ \mathcal{O}\Big(\frac{e^{\,\frac{1}{2}\,\im\frac{\sqrt{E}}{2}}}{E^2}\Big)\,.
\end{split}\end{equation}
Since $v$ has period 1 and mean 0, then
\begin{equation}\begin{split}
\mathcal{V}(\frac{1}{2})-\mathcal{V}(-\frac{1}{2}) & = \int_0^{1/2} \!\!\!v(\xi)\,\ud\xi - \int_0^{-1/2}\!\!\! v(\xi)\,\ud\xi \\
& = \int_{-1/2}^{1/2} v(\xi)\,\ud\xi = 0\,, \\
v(\frac{1}{2})+v(-\frac{1}{2}) & = 2\,v(\frac{1}{2}) \\
\mathcal{V}(-\frac{1}{2}) & = \int_0^{-1/2}\!\!\! v(\xi)\,\ud\xi \\
& = - \int_{-1/2}^{1/2}\! v(\xi)\,\ud\xi + \int_0^{1/2}\!\!\! v(\xi)\,\ud\xi = \mathcal{V}(\frac{1}{2})\,,
\end{split}\end{equation} 
whence
\begin{equation}\begin{split}
\mathcal{N}(E) & =\frac{2\sin\frac{\sqrt{E}}{2}}{\sqrt{E}}+\frac{\sin\frac{\sqrt{E}}{2}}{2 E^{3/2}}\,\Big[\,v(\frac{1}{2})+v(0)-\mathcal{V}^2(\frac{1}{2})\,\Big] \\ 
& \;\; + \mathcal{O}\Big(\frac{e^{\,\frac{1}{2}\,\im\frac{\sqrt{E}}{2}}}{E^2}\Big)\,.
\end{split}\end{equation}
Now, to square $\mathcal{N}(E)$ discarding the subleading terms recall  that
\begin{equation}\label{|sin|^2}
\Big|\,\sin\frac{\sqrt{E}}{2}\,\Big|\leqslant e^{\,\im\frac{\sqrt{E}}{2}}\,,\quad\Big|\,\cos\frac{\sqrt{E}}{2}\,\Big|\leqslant e^{\,\im\frac{\sqrt{E}}{2}}\qquad\forall\,E\in\mathbb{C}
\end{equation}
(see, e.g., (\ref{maggiorazioni_complesse}) in the appendix and the discussion thereafter) so that
\begin{equation}\begin{split}\label{N^2}
\mathcal{N}^2(E) & = \displaystyle\frac{4\sin^2\frac{\sqrt{E}}{2}}{E}+c\,\frac{\sin^2\frac{\sqrt{E}}{2}}{E^2}+\mathcal{O}\Big(\frac{e^{\im\sqrt{E}}}{E^{5/2}}\Big) \\
\mathrm{with}\;c & = 2\,\Big[\,v(\frac{1}{2})+v(0)-\Big(\int_0^{1/2}\!\!v(\xi)\,\ud\xi\Big)^2\,\Big]\,.
\end{split}\end{equation}

Analogously, from (\ref{2-D}) one gets
\begin{equation}\label{M^2}\begin{split}
\mathcal{M}^2(E) & = \frac{2-D(E)}{E-E_0} = \frac{\displaystyle 4\sin^2\frac{\sqrt{E}}{2}+ \Big(\frac{e^{\im\sqrt{E}}}{E^{3/2}}\Big)}{\displaystyle E\Big(1-\frac{E_0}{E}\Big)} \\
& = \left(\frac{4\sin^2\frac{\sqrt{E}}{2}}{E}+ \Big(\frac{e^{\im\sqrt{E}}}{E^{5/2}}\Big) \right)\cdot\left(1+\frac{E_0}{E}+\mathcal{O}\Big(\frac{1}{E^2}\Big)\right) \\
& = \displaystyle\frac{4\sin^2\frac{\sqrt{E}}{2}}{E}+4E_0\,\frac{\sin^2\frac{\sqrt{E}}{2}}{E^2}+\mathcal{O}\Big(\frac{e^{\im\sqrt{E}}}{E^{5/2}}\Big)
\end{split}\end{equation}
so that (\ref{step1}) is achieved and the first step is completed.

\medskip

\noindent\textsc{Step 2.} \emph{Pick any $\varepsilon>0$ small enough. Then $f^2$ is bounded in the closed region $\Omega_{\varepsilon}$ defined in} (\ref{omegaepsilon})\emph{, the bound depending on $\varepsilon$:}
\begin{equation}\label{Omega_boundedness}
\exists\,C_{\varepsilon}>0\,:\,|f^2(E)|\leqslant C_{\varepsilon}\quad\forall E\in\Omega_{\varepsilon}\,.
\end{equation}
\emph{In particular} (\ref{Omega_boundedness}) \emph{holds for every point on $\Gamma_{\varepsilon}$, the boundary of the angle $\mathbb{C}\setminus\Omega_{\varepsilon}$ with vertex at the origin, and}
\begin{equation}\label{limit_in_omega}
\lim_{E\to\infty} f^2(E) = 1\,, \qquad E\in\Omega_{\varepsilon}\,.
\end{equation}

Here a shorter truncation in (\ref{step1}) suffices, that is,
\begin{equation*}
\frac{\;\displaystyle\frac{4\sin^2\frac{\sqrt{E}}{2}}{E}+c\,\frac{\sin^2\frac{\sqrt{E}}{2}}{E^2}+\mathcal{O}\Big(\frac{e^{\im\sqrt{E}}}{E^{5/2}}\Big)}{\;\displaystyle\frac{4\sin^2\frac{\sqrt{E}}{2}}{E}+4E_0\,\frac{\sin^2\frac{\sqrt{E}}{2}}{E^2}+\mathcal{O}\Big(\frac{e^{\im\sqrt{E}}}{E^{5/2}}\Big)}=\frac{\;\displaystyle\frac{4\sin^2\frac{\sqrt{E}}{2}}{E}+\mathcal{O}\Big(\frac{e^{\im\sqrt{E}}}{E^2}\Big)}{\;\displaystyle\frac{4\sin^2\frac{\sqrt{E}}{2}}{E}+\mathcal{O}\Big(\frac{e^{\im\sqrt{E}}}{E^2}\Big)}\,.
\end{equation*}
Then
\begin{equation}\label{limit_of_f_in_omegaepsilon}
\lim_{\substack{E\to\infty \\E\in\Omega_{\varepsilon}}}f^2(E)=\lim_{\substack{E\to\infty \\E\in\Omega_{\varepsilon}}}\,\frac{\;\;\displaystyle\frac{4\sin^2\frac{\sqrt{E}}{2}}{E}+\mathcal{O}\Big(\frac{e^{\im\sqrt{E}}}{E^2}\Big)\;}{\;\;\displaystyle\frac{4\sin^2\frac{\sqrt{E}}{2}}{E}+\mathcal{O}\Big(\frac{e^{\im\sqrt{E}}}{E^2}\Big)\;}=1\,,
\end{equation}
because when $E\to\infty$ in $\Omega_{\varepsilon}$ the leading term both in the numerator and the denominator is
$E^{-1}\sin^2\frac{\sqrt{E}}{2}$
which dominates the $\mathcal{O}\Big(\frac{e^{\im\sqrt{E}}}{E^2}\Big)$--remainders.
Indeed, when $E\in\Omega_{\varepsilon}$ and $|E|$ is large enough, (\ref{|sin|^2}) actually becomes
\begin{equation}
\sin\frac{\sqrt{E}}{2}\sim e^{\im\frac{\sqrt{E}}{2}}
\end{equation}
(see, e.g., (\ref{maggiorazioni_complesse}) in the appendix and the discussion thereafter), and
\begin{equation}
\im\frac{\sqrt{E}}{2}=\frac{\sqrt{|E|}}{2}\sin\frac{\theta_E}{2}>\frac{\sqrt{|E|}}{2}\sin\frac{\varepsilon}{2}>0\,,\qquad E\in\Omega_{\varepsilon}\,.
\end{equation}
To conclude this step, by continuity the entire function $f$ can blow up neither at any finite point of $\Omega_{\varepsilon}$ nor at $\infty$ in $\Omega_{\varepsilon}$ and (\ref{Omega_boundedness}) is proved.

Notice that if 
an arbitrarily small angle around the positive real axis was not cut off, then $\im\frac{\sqrt{E}}{2}$ would not necessarily 
increase up to $+\infty$; furthermore the function $\sin^2\frac{\sqrt{E}}{2}$ has countably many zeroes 
in the real points $E_n=(2n\pi)^2$ and a priori it may not dominate the $\mathcal{O}\Big(\frac{e^{\im\sqrt{E}}}{E^2}\Big)$ remainder.

\medskip

\noindent\textsc{Step 3.} \emph{$f^2$ has a finite \emph{growth order} which does not exceed} $1$.

Recall (\cite{markushevich_chap}) that an entire function $h:\mathbb{C}\to\mathbb{C}$ is said to have finite growth order $\rho$ if 
\begin{equation}\label{finite_order}
\rho := \inf\left\{\mu>0\,:\,M(r)<e^{r^{\mu}}\right\}=\limsup_{r\to +\infty}\frac{\ln M(r)}{r^{\rho}}<+\infty
\end{equation}
where $M(r):=\max_{|z|=r}|h(z)|$. Otherwise $h$ is said to be of infinite order, and
\begin{equation}\label{infinite_order}
\lim_{z\to\infty}\frac{|h(z)|}{e^{|z|^{\rho}}}=+\infty\quad\forall\rho>0\,.
\end{equation}

We now see that our $f^2$ is not of infinite order. $\mathcal{M}(E)$ has only simple zeroes, that are all real;
since $f$ is analytic, these are also zeroes of $\mathcal{N}(E)$. From step 1 and $\forall\mu\in\mathbb{R}$,
\begin{equation*}
e^{-|E|^{\mu}}f^2(E)=\frac{\displaystyle\frac{4\sin^2\frac{\sqrt{E}}{2}}{e^{|E|^{\mu}}E}+\mathcal{O}\left(\frac{e^{\im\sqrt{E}}}{e^{|E|^{\mu}}E^2}\right)}{\displaystyle\frac{4\sin^2\frac{\sqrt{E}}{2}}{E}+\mathcal{O}\left(\frac{e^{\im\sqrt{E}}}{E^2}\right)}
\end{equation*}
is still an everywhere defined function, its numerator still vanishing whenever its denominator does; but now, due to (\ref{|sin|^2}), as long as $\mu\geqslant 1$
\begin{equation}\label{vanishing}
\frac{4\sin^2\frac{\sqrt{E}}{2}}{e^{|E|^{\mu}}E}\xrightarrow[E\to\infty]{}0\,,\quad \mathcal{O}\left(\frac{e^{\im\sqrt{E}}}{e^{|E|^{\mu}}E^2}\right)\xrightarrow[E\to\infty]{}0\,\qquad(\mu\geqslant 1)\,.
\end{equation}
This leads to
\begin{equation}
\frac{|f^2(E)|}{e^{|E|^{\mu}}}\xrightarrow[E\to\infty]{}0\,,
\end{equation}
namely (\ref{infinite_order}) fails to happen. That is, $f^2$ has finite growth order $\rho$. 
In view of (\ref{finite_order}), we are sure that $\rho$ is certainly dominated by the bound $\rho_\mathrm{b}=1$.

%
%

\medskip

\noindent\textsc{Step 4.} \emph{$f^2$ is bounded also on $\mathbb{C}\setminus\Omega_{\varepsilon}$, with the same bound as on $\Omega_{\varepsilon}$.}

In fact the following facts turned out to hold:
\begin{itemize}
\item[$a)$] $|f^2(E)|\leqslant C_{\varepsilon}\leq +\infty$ $\forall E\in\Gamma_{\varepsilon}$, the boundary of the angle $\mathbb{C}\setminus\Omega_{\varepsilon}$ (step 2);
\item[$b)$] $f^2$ is entire and its growth order does not exceed $\rho_\mathrm{b}=1$ (step 3);
\item[$c)$] the angle $\mathbb{C}\setminus\Omega_{\varepsilon}$ has arbitrarily small amplitude $2\varepsilon$, which can be taken less than $\displaystyle\frac{\pi}{\rho_\mathrm{b}}$, namely $\pi$.
\end{itemize}
As a consequence of a standard corollary of the Phragm\'en-Lindel\"of theorem (see, e.g., theorem 9.12 in \cite{markushevich_chap}), $a)$+$b)$+$c)$ imply that
\begin{equation}
|f^2(E)|\leqslant C_{\varepsilon}\qquad\forall E\in\mathbb{C}\setminus\Omega_{\varepsilon}\,.
\end{equation}

\medskip

\noindent\textsc{Conclusion of the proof.} Fixed a small enough angle with vertex at the origin, centred around the positive real axis, 
and with amplitude $2\varepsilon$, the entire function $f^2$ turns out to be bounded both inside this angle, namely on $\mathbb{C}\setminus\Omega_{\varepsilon}$, and outside of it, namely on $\Omega_{\varepsilon}$. That is, $f^2$ is bounded on the whole $\mathbb{C}$-plane, and so must be $f$ as well. Hence $f$ is identically constant, by the Liouville's theorem. Such a constant can be evaluated, e.g., taking the limit $E\to\infty$ in some suitable way, 
as in (\ref{limit_of_f_in_omegaepsilon}). So $f^2\equiv 1$. $\hspace{\stretch{1}}\square$


The crucial role of analyticity of $f^2$ is remarkable. In fact, before applying Phragm\'en-Lindel\"of theorem for analytic functions,
that is, without making use of analyticity, yet one has an apparently striking control on $f^2$: it vanishes
along any path $E\to\infty$ in $\mathbb{C}$ but an arbitrarily small angle centred at the origin around a given ray (the positive real axis).
However this does not suffice to claim boundedness on the whole $\mathbb{C}$. Neither it would suffice if in addition one knew
$f^2$ to vanish at infinity also along the positive real axis, or along any ray emanating from the origin. 
A paradigmatic counterexample for this phenomenon is the smooth $\mathbb{R}^2\to\mathbb{R}$ functions
\begin{equation}
h(x,y) = (x^2+y^2)\,e^{-(x-y^{2})^2}\,.
\end{equation}
Indeed for any $\varepsilon>0$ small enough
\begin{equation*}\begin{split}
\lim_{x\to +\infty} h(x,0) & =0 \\
\lim_{(x,y)\to\infty}h(x,y) & =0 \quad
\begin{array}{l}\forall (x,y)\in\mathbb{R}^2\;\;\mathrm{such}\;\mathrm{that} \\
\theta :=\arg(x,y)\in [\,\varepsilon,2\pi-\varepsilon\,] 
\end{array}
\end{split}\end{equation*}
because, when $\theta\in [\,\varepsilon,2\pi-\varepsilon\,]$ (so that $|\sin\theta|\geqslant\sin\varepsilon >0$) and $r:=\sqrt{x^2+y^2}$ is large enough,
\begin{equation*}
(x-y^{2})^2 = r^{4}(\sin^2\theta - r^{-1}\cos\theta)^2 \geqslant\frac{r^{4}\sin^4\varepsilon}{2}
\end{equation*}
so
\begin{equation*}
0\leqslant h(x,y)\leqslant r^2 \,e^{-\frac{1}{2}r^{4}\sin^4\varepsilon}\xrightarrow[r\to +\infty]{}0\,.
\end{equation*}
Yet $h$ diverges as $r^2$ on the curve $x-y^{2}=0$.

\section*{Acknowledgements}
We are indebted to S.~Baroni and G.~Santoro for motivating the original interest to this topic.
Helpful and critical discussions with R.~Adami, V.~Carnevale, G.~Dell'An\-to\-nio, B.~A.~Dubrovin and G.~Morchio are warmly acknowledged.

  \renewcommand{\theequation}{A.\arabic{equation}}
  \setcounter{equation}{0}  

\section*{Appendix. \\ Spectral analysis of 1D periodic potentials \\and connections with Hill's equation theory}

\subsection*{A.1 -- Spectral analysis. Degeneracy.}

As a starting point \cite{rs4_in,gp2000_in} one has to recognise that the space of wave functions on which the Hamiltonian
(\ref{Hamiltonian}) acts decomposes naturally into subspaces labelled by the boundary conditions at $x=0$ and $x=1$.
This way one has to consider the one-particle Hamiltonians
\begin{equation}
H(k) = \Big(-\frac{\ud^2}{\ud x^2}\Big)_{k}+v(x)\label{H(k)}
\end{equation}
acting on the (dense) domain of self-adjointness $\mathcal{D}_k$ of the measurable functions on $[0,1]$ that are
square-summable together with their first two derivatives, and satisfy the boundary conditions
\begin{equation}\label{k-periodicity}
\psi(1)=e^{ik}\psi(0)\quad,\quad\psi'(1)=e^{ik}\psi'(0)\,.
\end{equation}
(For convenience we rename $\mathcal{D}_+:=\mathcal{D}_0$ and $\mathcal{D}_-:=\mathcal{D}_{\pi}$.) Thus the quantum-mechanical
problem on $\mathbb{R}$ is rephrased in terms of the problem of a Schr\"odinger-like particle on $[0,1]$ with boundary conditions
labelled by $k$.

The spectral analysis of $H$ on the whole is the union of the spectral analyses of the $H(k)$'s, and the following facts hold:
\begin{itemize}
\item each $H(k)$ has a purely discrete spectrum;
\item its eigenvalues $E_1(k)$, $E_2(k)$, $E_3(k)$, $\dots$ are nondegenerate for any $k\in(0,\pi)$; 
\item $E_1(0)$ is nondegenerate as well;
\item $k\mapsto E_n(k)$ is analytic in $(0,\pi)$ and continuous in $[0,\pi]$;
\item for $k\in[0,\pi]$, $k\mapsto E_n(k)$ is monotone increasing for $n$ odd and monotone decreasing for $n$ even, and 
\begin{equation*}
E_1(0)<E_1(\pi)\leq E_2(\pi)<E_2(0)\leq E_3(0)<E_3(\pi)\leq\cdots\,;
\end{equation*}
\item $H(k)$ and $H(-k)$ are antiunitarily equivalent under ordinary complex conjugation, in particular their eigenvalues are identical and their eigenfunctions are complex conjugates.
\end{itemize}

The \emph{dispersion relations} $k\mapsto E_n(k)$ produce the familiar structure with bands and gaps (Fig.~\ref{bande}).
$E_n(\cdot)$ is the $n$-th band within the Brillouin zone. Also, any $\psi_k\in\mathcal{D}_k$, when extended on $\mathbb{R}$
with boundary conditions like (\ref{k-periodicity}) on every interval $[n,n+1]$, is the Bloch function
$\psi_k(x)=e^{ikx}u_k(x)$ with $u_k(x+1)=u_k(x)$. If
\begin{equation}
\alpha_n\!:=\left\{
\begin{array}{ll}
E_n(0) & n\;\mathrm{odd} \\
E_n(\pi) & n\;\mathrm{even}
\end{array},\right.\;\;
\beta_n\! :=\left\{
\begin{array}{ll}
E_n(\pi) & n\;\mathrm{odd} \\
E_n(0) & n\;\mathrm{even}
\end{array}\right.\,,
\end{equation}
then $[\alpha_n,\beta_n]$ is the $n$-th band, $(\beta_n,\alpha_{n+1})$ is the $n$-th gap, i.e., the gap between $n$-th and the $(n+1)$-th band, and $H$ has a purely absolutely continuous spectrum $\sigma(H)=\bigcup_{n=1}^{\infty}[\alpha_n,\beta_n]$.

Eigenfunctions at the band edge $k=0$ have period equal to $1$, satisfying the periodic boundary conditions. Eigenfunctions at the band edge $k=\pi$ have period equal to $2$, due to the antiperiodic boundary conditions.

With this notation, the $n$-th gap vanishes iff $\beta_n=\alpha_{n+1}$: this equivalently corresponds to $E_n(0)=E_{n+1}(0)$ when $n$ is even, and to $E_n(\pi)=E_{n+1}(\pi)$, 
when $n$ is odd. Consequently one band and the following one collapse into a unique band at the centre of the Brillouin zone at the energy $E$, iff $E$ is a \emph{doubly degenerate} eigenvalue of $H(0)$. Similarly they collapse at the edge of the Brillouin zone iff $E$ is a \emph{doubly degenerate} eigenvalue of $H(\pi)$.

\subsection*{A.2 -- Hill's equation theory. Main features.}

The dispersion relations $k\mapsto E_n(k)$ can be understood as branches of the parametrisation \cite{gp2000_in,ash_chap}
\begin{equation}\label{cosk}
\frac{\cos\big(\sqrt{E}+\delta(E)\big)}{|\,t(E)\,|}=\cos k
\end{equation}
of the variable $E$ in terms of the parameter $k$, where $\delta(E)$ and $|\,t(E)\,|$ are the phase and the modulus of the
transmission amplitude $t(E)$ of a free particle (i.e., a plane wave) of energy $E$ incident on a single cell of the lattice
described by $v$. Twice the left hand side of (\ref{cosk}) is a quantity known as
the \emph{discriminant} $D(E)$ of Hill's differential equation
\begin{equation}\label{ourHill}
-\psi''+v\,\psi=E\psi\,
\end{equation}
and this bridges the spectral analysis of 1D periodic potentials to the Hill's equation theory \cite{magnus,eastham,mckean75,trubowitz77,gt84}.

Such a discriminant, by definition, is
\begin{equation}\label{DISCRIMINANT}
D(E) :=\psi_1^{[E]}(1)+(\psi_2^{[E]})'(1)
\end{equation}
where $E$ is now allowed to be complex, and $\psi_1^{[E]}(x)$ and $\psi_2^{[E]}(x)$ are the so called
\emph{fundamental solutions} of the Hill's equation, that is, by definition, the solutions of the Cauchy problems
\begin{equation}\label{fundamentalsoll}
\begin{split}
&\left\{
\begin{array}{rll}
-(\psi_1^{[E]})''+v\,\psi_1^{[E]}&=&E\psi_1^{[E]} \\
\psi_1^{[E]}(0)&=&1 \\
(\psi_1^{[E]})'(0)&=&0
\end{array}\right.\;\; \\
& \\
& \left\{
\begin{array}{rll}
-(\psi_2^{[E]})''+v\,\psi_2^{[E]}&=&E\psi_2^{[E]} \\
\psi_2^{[E]}(0)&=&0 \\
(\psi_2^{[E]})'(0)&=&1
\end{array}\right.
\end{split}
\end{equation}

The discriminant turns out to be an entire function of the complex variable $E$ with some crucial properties in dealing
with the general solutions of (\ref{ourHill}).
Within this O.D.E. framework, what one commonly refers to as \emph{energy band}, \emph{gap}, \emph{band edge eigenvalues},
\emph{vanishing gaps}, translate respectively into the concepts of \emph{interval of stability}, \emph{interval of
instability}, \emph{characteristic values}, \emph{coexistence}, with the features we sketch here below.

\begin{figure}
\begin{center}
\includegraphics[height=5cm]{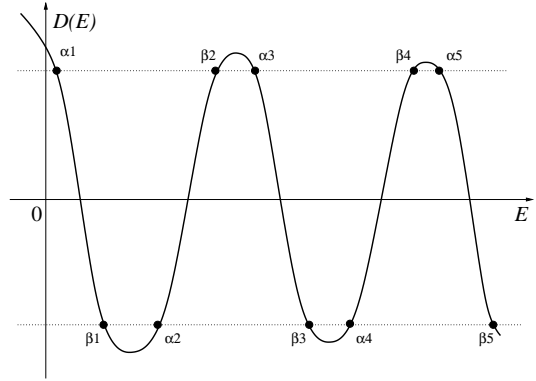}
\end{center}
\caption{Discriminant behaviour}\label{discr}
\end{figure}

\begin{itemize}
\item For real $E$'s, $D(E)$ has a graph somewhat like that in Fig.~\ref{discr}. In fact, the following asymptotic behaviour is known:
\begin{equation}\label{D-behaviour}
\begin{array}{lll}
D(E) = 2\cos\sqrt{E} + \mathcal{O}(E^{-3/2}) & & E\in\mathbb{R},\; E\to +\infty \\
D(E) = 2\cosh\sqrt{|E|}\,\cdot\,(1+o(1)) & & E\in\mathbb{R},\; E\to -\infty
\end{array}
\end{equation}
For complex $E$'s, $D(E)$ is an entire function of growth order $\frac{1}{2}$ and type $1$ (\cite{markushevich_chap}).
\item For any complex $E$, or real $E$ such that $|D(E)|>2$, then all nontrivial solutions of (\ref{ourHill}) are unbounded
in $(-\infty,+\infty)$; unbounded solutions are called \emph{unstable}. If $E$ is real and $|D(E)|<2$ then all nontrivial
solutions of (\ref{ourHill}) are bounded in $(-\infty,+\infty)$; bounded solutions are called \emph{stable}.
\item $D(E)=2$ has infinitely many real roots, which are single or at most double, increasing up to infinity, denoted in increasing order by
\[
\alpha_1,\beta_2,\alpha_3,\beta_4,\alpha_5,\beta_6,\dots\longrightarrow +\infty
\]
and $D(E)=-2$ has infinitely many real roots, which are single or at most double, increasing up to infinity, denoted in increasing order by
\[
\beta_1,\alpha_2,\beta_3,\alpha_4,\beta_5,\alpha_6,\dots\longrightarrow +\infty 
\]
so that whenever $E$ belongs to any of the so-called \emph{stability intervals} $(\alpha_n,\beta_n)$, then all the corresponding solutions of (\ref{ourHill}) are stable, whereas if $E$ belongs to any of the so-called \emph{instability intervals} $(\beta_n,\alpha_{n+1})$, or even $(-\infty,\alpha_1)$, then all the corresponding solutions of (\ref{ourHill}) are unstable.
\item The $n$-th stability interval has a width $(\beta_n-\alpha_n)\lesssim(2n-1)\pi^2$ and $(\beta_n-\alpha_n)-(2n-1)\pi^2\to 0$ as $n\to\infty$. The asymptotic behaviour of the $n$-th instability interval width is $(\alpha_{n+1}-\beta_n)\to 0$ and the positions of both edges $\alpha_{n+1}$, $\beta_n$ are asymptotically given by $n^2\pi^2+\int_0^1 v(x)\ud x$ .
\item All the solutions of (\ref{ourHill}) satisfy the boundary conditions (\ref{k-periodicity})
\begin{equation*}
\quad\psi(x+1)=e^{ikx}\psi(x)\;\;,\;\;\psi'(x+1)=e^{ikx}\psi'(x)
\end{equation*}
or, \emph{equivalently}, $E\in\mathbb{R}$ is an eigenvalue of $H(k)$, if and only if $D(E)=2\cos k$. In particular solutions $\psi$ with $D(E)=2$ are 1-periodic, $\psi$'s with $D(E)=-2$ are 2-periodic. This gives the correspondence between the bands/gaps and the stability/instability intervals.
\item Whenever $D(E)=2$ has a \emph{double} real root, say $\widetilde{E}=\beta_{2n}=\alpha_{2n+1}$, then the $2n$-th interval of instability $(\beta_{2n},\alpha_{2n+1})$ disappears and there coexist two linearly independent 1-periodic solutions of (\ref{ourHill}) with $E=\widetilde{E}$ (and therefore all the solutions are 1-periodic at that $\widetilde{E}$). Analogously, \emph{coexistence} happens whenever $D(E)=-2$ has a \emph{double} real root $\widetilde{E}=\beta_{2n+1}=\alpha_{2n+2}$: there are two linearly independent 2-periodic solutions of (\ref{ourHill}) with that $E=\widetilde{E}$.
\end{itemize}

Thus one sees that the vanishing gap phenomenon translates into the coexistence phenomenon for Hill's equation, associated to the double zeroes of $2-D(E)=0$ or $2+D(E)=0$.

\subsection*{A.3 -- Asymptotic estimates}

We conclude this appendix discussing the asymptotic expansions as $E\to\infty$ of the fundamental solutions (\ref{fundamentalsoll}).
Actually what we are doing in the following is to restate some classical but heterogeneous material \cite{mckean76,magnus,ungar61,hoch63,hoch65,h77,gt84} in an organic unified perspective.

The integral representation of the fundamental solutions is
\begin{equation}\label{integral_representation}\!\!\begin{array}{rll}
\psi_1^{[E]}(x) & = & \cos x\sqrt{E} + \displaystyle\int_0^x\frac{\sin[(x-\xi)\sqrt{E}]}{\sqrt{E}}\,v(\xi)\,\psi_1^{[E]}(\xi)\,\ud\xi \\
\psi_2^{[E]}(x) & = & \displaystyle\frac{\sin x\sqrt{E}}{\sqrt{E}}+\displaystyle\int_0^x\frac{\sin[(x-\xi)\sqrt{E}]}{\sqrt{E}}\,v(\xi)\,\psi_2^{[E]}(\xi)\,\ud\xi
\end{array}\end{equation}
(these identities can be derived in a swift manner by Laplace transformation of (\ref{fundamentalsoll})), whence, by iteration,
\begin{equation}\label{expansions}\begin{split}
\psi_1^{[E]}(x) &= \cos x\sqrt{E} \\
& +\sum_{k=1}^{N-1}\left[P_k^{[1]}(x)\frac{\sin x\sqrt{E}}{E^{k-\frac{1}{2}}}+Q_k^{[1]}(x)\frac{\cos x\sqrt{E}}{E^{k}}\right] \\
& +\mathcal{O}\left(\frac{\sin x\sqrt{E}}{E^{N-\frac{1}{2}}}\right) + \mathcal{O}\left(\frac{\cos x\sqrt{E}}{E^{N}}\right) \\
\psi_2^{[E]}(x) &= \frac{\sin x\sqrt{E}}{\sqrt{E}} \\
& +\sum_{k=1}^{N-1}\left[Q_k^{[2]}(x)\frac{\cos x\sqrt{E}}{E^{k}}+P_{k+1}^{[2]}(x)\frac{\sin x\sqrt{E}}{E^{k+\frac{1}{2}}} \right] \\
& + \mathcal{O}\left( \frac{\cos x\sqrt{E}}{E^{N}}\right) + \mathcal{O}\left(\frac{\sin x\sqrt{E}}{E^{N+\frac{1}{2}}}\right)\,.
\end{split}\end{equation}
Here $P_k^{[i]}$, $Q_k^{[i]}:[0,1]\to\mathbb{R}$ ($i=1,2$; $k=1,\dots ,N-1$) are the coefficients of
$\frac{\sin x\sqrt{E}}{E^{k-\frac{1}{2}}}$ and $\frac{\cos x\sqrt{E}}{E^{k}}$ respectively in the expansion of the $i$-th fundamental solution.
They are suitable continuous (and hence bounded) functions of $x\in [\,0,1]$,
independent of the parameter $E$, and depending only on the choice of the potential $v$. 
Their existence is guaranteed provided $v$ is sufficiently differentiable.
In  particular $P_1^{[2]}(x)\equiv 1$.

For instance, for real positive values of the parameter $E$, and if $v$ is assumed to be only continuous (as in our main theorem), one finds
\begin{equation}\label{estimates_R_real}{}\!\!\!\!\!\!\!\!\!\begin{array}{lll}
\psi_1(x,E) &=& \cos x\sqrt{E}+\displaystyle\frac{\sin x\sqrt{E}}{2\sqrt{E}}\mathcal{V}(x) \\
& & +\displaystyle\frac{\cos x\sqrt{E}}{4E}[v(x)-v(0)-\frac{1}{2}\mathcal{V}^2(x)]+\mathcal{O}\Big(\frac{1}{E^{3/2}}\Big) \\
& & \\
\psi_2(x,E) &=& \displaystyle\frac{\sin x\sqrt{E}}{\sqrt{E}}-\displaystyle\frac{\cos x\sqrt{E}}{2E}\mathcal{V}(x) \\
& & +\displaystyle\frac{\sin x\sqrt{E}}{4E^{3/2}}[v(x)+v(0)-\frac{1}{2}\mathcal{V}^2(x)]+\mathcal{O}\Big(\frac{1}{E^{2}}\Big)
\end{array}\end{equation}
with 
\begin{equation}
\mathcal{V}(x):=\int_0^x v(\xi)\,\ud\xi\,. 
\end{equation}
The form of the remainders in (\ref{estimates_R_real}) is just a shortcut to include two different types of remainders appearing in (\ref{expansions}):
both $\sin x\sqrt{E}$ and $\cos x\sqrt{E}$ do not exceed 1, since $E\in\mathbb{R}_+$.
Incidentally, notice that plugging (\ref{estimates_R_real}) into the definition
(\ref{DISCRIMINANT}) of the discriminant, immediately gives the first equation in (\ref{D-behaviour}). 

The above expansions hold when $E\to\infty$ in the \emph{complex} plane. 
The only difference with respect to the special real case is that in general the $N$-th remainders can diverge (in modulus) as $E\to\infty$, as remarked
in Section \ref{proof_of_lemma}.

From (\ref{expansions}) one deduces an analogous estimate for the quantity $D(E)$, 
Indeed, plugging (\ref{expansions}) into (\ref{DISCRIMINANT}) and after some lengthy manipulations one gets
\begin{equation}\label{D-expanded}\begin{split}
2&-D(E) = 4\sin^2\frac{\sqrt{E}}{2} \\
& \;\;+ \sum_{k=1}^{N-1}\left[\alpha_k \frac{\sin \frac{\sqrt{E}}{2} \cos \frac{\sqrt{E}}{2} }{E^{k-\frac{1}{2}}} +\beta_k\frac{\sin^2 \frac{\sqrt{E}}{2}}{E^{k}} + \gamma_k\frac{1}{E^k}\right] \\
& \;\;+\mathcal{O}\left( \frac{\sin \frac{\sqrt{E}}{2} \cos \frac{\sqrt{E}}{2} }{E^{N-\frac{1}{2}}}  \right)   +\mathcal{O}\left(\frac{\sin^2 \frac{\sqrt{E}}{2}}{E^{N}} \right)   +\mathcal{O}\left(\frac{1}{E^N} \right)
\end{split}\end{equation}
with
\begin{equation}\begin{split}
\alpha_k & = -2\left[\,P_k^{[1]}(1)+(P_k^{[2]})'(1)-Q_k^{[2]}(1)\,\right] \\
\beta_k   = -2\gamma_k & = -2 \left[\,Q_k^{[1]}(1)+(Q_k^{[1]})'(1)+P_{k+1}^{[2]}(1)\,\right]\,.
\end{split}\end{equation}

As for the particular case of real $E$'s, it is customary to condense all the remainders of (\ref{expansions}) and of (\ref{D-expanded})  into a unique form.
This is done by noticing that $\forall\,\alpha\in\mathbb{C}$
\begin{equation}\label{maggiorazioni_complesse}\begin{split}
|\sin\alpha\,|^2&=\left|\frac{e^{i\alpha}-e^{-i\alpha}}{2}\right|^2
\\
&= \frac{1}{2}\left(\cosh\im(2\alpha)-\cos\re(2\alpha) \right) \leqslant e^{\,|\im(2\alpha)|} \\
|\cos\alpha\,|^2 &= \frac{1}{2}\left(\cosh\im(2\alpha)+\cos\re(2\alpha) \right) \leqslant e^{\,|\im(2\alpha)|}
\end{split}\end{equation}
Thus, remainders can be written as $\mathcal{O}(e^{|\,x\,\im\frac{\sqrt{E}}{2}\,|}E^{-n})$ and $\mathcal{O}(e^{|\,\im\sqrt{E}\,|}E^{-n})$ respectively.
Also, the absolute value can be removed since $x\in [\,0,1]$ and $\im\sqrt{E}\geqslant 0$, due to the standard determination of the square root in the complex plane
(if $\mathbb{C}\ni E=|E|e^{\,\mathrm{arg}(E)}$ with $0\leqslant\mathrm{arg}(E)\leqslant 2\pi$, then
$\sqrt{E}=\sqrt{|E|}\,e^{\,\mathrm{arg}(E)/2}$ with $0\leqslant\mathrm{arg}(E)/2\leqslant \pi$, whence $\im\sqrt{E}\geqslant 0$).

\emph{In the case of zero-mean potential}, when suitably truncating (\ref{expansions}) and (\ref{D-expanded}) one gets exactly the expansions (\ref{2-D}) and (\ref{psi2}) used in our proof. As a further example, going to higher orders one would get
\begin{equation}\begin{split}\label{2-D-extended}
2-D(E)&= 4\sin^2\frac{\sqrt{E}}{2} \,-\,\frac{\sin\frac{\sqrt{E}}{2}\cos\frac{\sqrt{E}}{2}     }{2 E^{3/2}}\int_0^1 \!v^2(\xi)\,\ud\xi \\
&- \frac{\sin\frac{\sqrt{E}}{2}\cos\frac{\sqrt{E}}{2}}{8 E^{5/2}}\int_0^1 \!\left[2v^3(\xi)+v'(x)^2\right]\,\ud\xi \\
&+\mathcal{O}\Big(\frac{e^{\im\sqrt{E}}}{E^{3}}\Big)\,
\end{split}\end{equation}
\begin{equation}\begin{split}\label{psi2-extended}
\psi_2^{[E]}(x)&=\frac{\sin x\sqrt{E}}{\sqrt{E}}-\frac{\cos x\sqrt{E}}{2E}\mathcal{V}(x)\\
&+\frac{\sin x\sqrt{E}}{4 E^{3/2}}\,\Big[\,v(x)+v(0)-\frac{1}{2} \mathcal{V}(x)^2\,\Big] \\
&+ Q_2^{[2]}(x)\cdot\frac{\cos x\sqrt{E}}{E^{2}_{\;}} + P_3^{[2]}(x)\cdot\frac{\sin x\sqrt{E}}{E^{5/2}}\\
&+\mathcal{O}\Big(\frac{e^{\,x\,\im\frac{\sqrt{E}}{2}}}{E^3}\Big)
\end{split}\end{equation}
where the coefficients $Q_2^{[2]}$ and $P_3^{[2]}$ (according to the notation in (\ref{expansions})\,) are given by
\begin{equation*}\begin{split}
Q_2^{[2]}(x) &= \displaystyle\frac{1}{8}\Big\{\frac{\mathcal{V}^3(x)}{3!} \\
&\;\; -\int_0^x v^2(\xi)\,\ud\xi - \mathcal{V}(x)\,\big[\,v(x)+v(0)\,\big]+v'(x)-v'(0)\Big\}\\
P_3^{[2]}(x) &= \displaystyle\frac{\mathcal{V}^4(x)}{4!}-\mathcal{V}(x)\int_0^x v^2(\xi)\,\ud\xi + \frac{5}{2}\,\big[\,v^2(x)+v^2(0)\,\big] \\
&\;\;+v(x)v(0) -\displaystyle v''(x)+v''(0)+\mathcal{V}(x)\,\big[\,v'(x)-v'(0)\,\big] \\
& \;\; -\frac{\mathcal{V}^2(x)}{4}\,\big[\,v(x)+v(0)\,\big]\,.
\end{split}\end{equation*}
Notice that higher order derivatives of $v$ occur.

\bibliographystyle{acm}
\bibliography{bib_iff}

\vspace{2cm}
Authors' address:
\medskip

International School of Advanced Studies (S.I.S.S.A.)

via Beirut 4, 34014 Trieste (TS)

Italy

\medskip

E-mail: osvaldo.zagordi@gmail.com, alemiche@sissa.it

\end{document}